\documentclass[aps,prd,twocolumn,showpacs,superscriptaddress,preprintnumbers,floatfix,reprint]{revtex4}

\usepackage{graphicx}
\usepackage{dcolumn}
\usepackage{bm}
\usepackage{yhmath}
\usepackage{hyperref}
\usepackage{amsmath}
\usepackage{booktabs}
\usepackage{subfigure}
\usepackage{xcolor}

\usepackage{multirow}
\usepackage{dcolumn}
\newcolumntype{C}[1]{>{\centering\arraybackslash}p{#1}}

\begin{document}

\preprint{
\vbox{
\hbox{ADP-15-46/T948}
}}

\def\Tr{{\rm Tr}}
\newcommand{\sla}{\not\!}
\def\sca{0.23}

\title{Hamiltonian effective field theory study of the $\mathbf{N^*(1535)}$ resonance in lattice QCD}

\author{Zhan-Wei Liu}
\affiliation{Special Research Center for the Subatomic Structure of Matter (CSSM), Department of Physics, University of Adelaide, Adelaide SA 5005, Australia}

\author{Waseem Kamleh}
\affiliation{Special Research Center for the Subatomic Structure of Matter (CSSM), Department of Physics, University of Adelaide, Adelaide SA 5005, Australia}

\author{Derek B. Leinweber}
\affiliation{Special Research Center for the Subatomic Structure of Matter (CSSM), Department of Physics, University of Adelaide, Adelaide SA 5005, Australia}

\author{Finn M. Stokes}
\affiliation{Special Research Center for the Subatomic Structure of Matter (CSSM), Department of Physics, University of Adelaide, Adelaide SA 5005, Australia}

\author{Anthony W. Thomas}
\affiliation{Special Research Center for the Subatomic Structure of Matter (CSSM), Department of Physics, University of Adelaide, Adelaide SA 5005, Australia}\affiliation{ARC Centre of Excellence in Particle Physics at the Terascale, Department of Physics, University of Adelaide, Adelaide SA 5005, Australia}

\author{Jia-Jun Wu}
\affiliation{Special Research Center for the Subatomic Structure of Matter (CSSM), Department of Physics, University of Adelaide, Adelaide SA 5005, Australia}

\begin{abstract}
Drawing on experimental data for baryon resonances, Hamiltonian
effective field theory (HEFT) is used to predict the positions of the
finite-volume energy levels to be observed in lattice QCD simulations
of the lowest-lying $J^P=1/2^-$ nucleon excitation.  In the initial
analysis, the phenomenological parameters of the Hamiltonian model are
constrained by experiment and the finite-volume eigenstate energies
are a prediction of the model.  The agreement between HEFT predictions
and lattice QCD results obtained on volumes with spatial lengths of 2
and 3 fm is excellent. These lattice results also admit a more
conventional analysis where the low-energy coefficients are
constrained by lattice QCD results, enabling a determination of
resonance properties from lattice QCD itself.  Finally, the role and
importance of various components of the Hamiltonian model are
examined.
\end{abstract}

\pacs{14.20.Gk, 
      12.38.Gc, 
      13.75.Gx  
}

\keywords{Baryon resonances, Effective Field Theory, Lattice QCD,
  Finite-Volume, Nucleon spectrum, Odd parity}

\maketitle

\pagenumbering{arabic}

Lattice QCD has proven remarkably successful in reproducing the masses and 
many other properties of the octet baryons, which are stable under the 
strong interaction. In our on-going quest to understand the structure 
of hadronic systems in terms of QCD, the focus is now shifting to 
excited states. Perhaps the greatest challenge there is that all states 
studied on a Euclidean space-time lattice are stable eigenstates of the QCD 
Hamiltonian, subject to periodic spatial boundary conditions. In 
contrast, the resonant states revealed in experiments are neither 
stable, nor are they eigenstates of the QCD Hamiltonian. Rather, they 
are often extremely short-lived, with multiple decay modes. Clearly one 
faces an enormous challenge when one aims to use lattice QCD to study 
these states.

One powerful technique, introduced by 
L\"uscher~\cite{Luscher:1985dn,Luscher:1986pf}, 
which has been widely used 
by the community, does provide a robust link between the discrete energy 
levels observed in lattice QCD and the scattering phase shifts 
extracted from experiment. This method presents technical complications 
when the resonance under study can decay through more than one open channel.
These complications can be overcome and the resulting 
formalism has been successfully applied in the coupled $\pi \pi$ 
and $K \bar{K}$ system~\cite{Wu:2014vma}. On the other hand,
several groups have been led to explore an 
alternative approach, which we label Hamiltonian effective field theory 
(HEFT).

HEFT enables a
quantitative examination of experimental observations such as
resonance positions, partial decay widths, scattering phase shifts and
inelasticities in terms of a model built from hadronic
degrees of freedom and their interactions.
While formulated in infinite volume, such models have
recently been applied to the analysis of the hadronic excitation
spectra observed in a small number of finite volume lattice QCD 
calculations~\cite{Hall:2013qba,Hall:2014uca}, namely   
the $\Delta$ resonance~\cite{Hall:2013qba} and 
the $\Lambda(1405)$~\cite{Hall:2014uca}. 
The former is 
a classical case where a three-quark state is dressed by coupling to 
the open $\pi N$ channel, while the latter is far more complex and 
illustrates some of the power of HEFT. In concert with a lattice study 
of the individual quark flavor 
contributions to the magnetic form factor of the 
baryon, the application of HEFT led to a deeper understanding of 
the nature of this resonance which has been mysterious for 50 years. 
That study strongly suggested that the $\Lambda(1405)$ does not 
have a significant three-quark component in its wave function, rather 
it is appropriately viewed as a $\bar{K} N$ bound state.

In this Letter we examine the nature of the first negative parity 
excitation of the nucleon, the $J^P = 1/2^- \, N^*(1535)$. This state has 
been the subject of much speculation in the 
literature~\cite{Nieves2001,Kaiser1995,Inoue2002,Liu2006}, since it lies 
above the first positive parity nucleon excited state (the Roper resonance 
at 1440 MeV), unlike the expectation in the phenomenologically 
very successful harmonic oscillator model. There have also been suggestions 
that there may be a significant strange quark component in this resonance, 
so it could be viewed as a penta-quark. Such questions are central to the 
modern study of resonances and with its $S$-wave coupling to both 
$\pi N$ and $\eta N$ channels this is an ideal case for study using 
HEFT to analyse modern lattice data. 
Our study supports the interpretation of the $N^*(1535)$
 as primarily a three-quark excitation, with couplings to five-quark 
components. The states most likely associated with the resonance
have a probability of about 50\% to contain the bare baryon, at 
the physical pion mass, in boxes with $L\simeq2, 3$ fm.

The HEFT used here introduces a bare state, $N^*_0$, 
which may be thought of as a three-quark 
state that would be stable in the absence of coupling to the
$\pi N$ and $\eta N$ channels. 
We do not consider the corrections from 
$N\pi\pi$ states, which would add significant technical complications, 
because the branching ratio in the case of the $N(1535)$ 
is only a few percent~\cite{Agashe:2014kda}.
The corresponding Hamiltonian has two parts, a non-interacting or bare
Hamiltonian, $H_0$, and an interacting Hamiltonian, $H_I$. 
In the center-of-mass system, the non-interacting part is
\begin{eqnarray}
H_0 &=& 
|N^*_0\rangle\,  m_{0} \, \langle N^*_0| + 
\sum_{\alpha} \int d\vec{k}\nonumber\\
&&  |\alpha(\vec{k})\rangle\, \omega_\alpha(k)\,
\langle\alpha(\vec{k})| \, .
\label{eq:h0}
\end{eqnarray}
Here $m_{0}$ is the mass of $N^*_0$, 
while $|\alpha(\vec{k})\rangle$ denotes either the $\pi N$ or $\eta N$
channel and $\omega_\alpha(k)$ is the corresponding energy, 
$\omega_\alpha(k)=\sqrt{m_{\alpha_1}^2 + \vec{k}^{2}} +
\sqrt{m_{\alpha_2}^2 + \vec{k}^{2}}$.  
Here $m_{\alpha_1}$ and $m_{\alpha_2}$ are masses of the meson and
baryon, respectively.

Following Refs.~\cite{Kamano2013,Matsuyama2007,Kamano2011}
where there was an extensive study of scattering data 
involving nucleon resonances up to 1.8 GeV, 
the interaction Hamiltonian can be divided into two parts, 
$H_I = g + v$.  
Here $g$ describes the interaction between the bare state
$N^*_0$ and the multi-particle channels which dress it:
\begin{eqnarray}
g = \sum_{\alpha}\, \int d\vec{k}\, \left \{
|\alpha(\vec{k})\rangle\, G^\dagger_{\alpha}(k)\,  \langle N^*_0| +
|N^*_0\rangle\, G_{\alpha}(k)\, \langle \alpha(\vec{k})| \right \} .
\label{eq:int-g}
\end{eqnarray}
Here, we take $G_{i  N}^2(k)=\frac{3\, g_{N^*_0 i  N}^2 }{4\pi^2\, f^2}\, 
\omega_i(k)\,  u^2(k)$, with $i= \pi$ or $\eta$ and 
$\omega_X(k)=\sqrt{k^2+m_X^2}$.
The pion decay constant is $f=92.4$ MeV and 
the regulator is taken to be a dipole with mass parameter 
$\Lambda=0.8$ GeV. $G_\alpha(k)$ corresponds to 
the Lagrangian $i\bar N^*_0 \gamma^\mu\partial_\mu \pi N+h.c.$, 
in the limit where the baryons are treated non-relativistically
and a dipole regulator is used to render the theory finite.

The second part of the interaction Hamiltonian 
is purely phenomenological. It is taken to be separable, with form factors
chosen to reproduce the low energy scattering data, well below the 
energy region where the resonance dominates.
It describes the
transitions between meson-baryon state $|\alpha(\vec{k})\rangle$ and
meson-baryon state $|\beta(\vec{k'})\rangle$:
\begin{eqnarray}
v = \sum_{\alpha,\beta}\, \int d\vec{k}\, d\vec{k}'\,
|\alpha(\vec{k})\rangle\,  V^{S}_{\alpha,\beta}(k,k')\, \langle
\beta(\vec{k}')| \, . 
\label{eq:int-v}
\end{eqnarray}
For example, the separable form for the interaction 
in the  $\pi N$ channel is
\begin{equation}
V_{\pi N, \pi N}^S(k,k') = \frac{3\, g^S_{\pi N}\, \tilde{u}(k)\,
  \tilde{u}(k')}{4\pi^2\, f^2} \, .
\label{eqVspiN}
\end{equation}
In order to fit the low energy experimental data well it was found that 
the form factors needed enhancement at low momentum and the purely 
ad hoc form $\tilde{u}(k) = u(k) (m_\pi+\omega_\pi(k))/\omega_\pi(k)$, 
works very well, with $u(k)$ the same function 
used in Eq.~(\ref{eq:int-g}). Of course, 
when exploring the fit to lattice data away from the physical pion mass,  
the value of $m_\pi$ appearing in $\tilde{u}$ is {\em not} varied.

\begin{figure*}[t]
\begin{center}
\subfigure{\scalebox{0.45}{\includegraphics{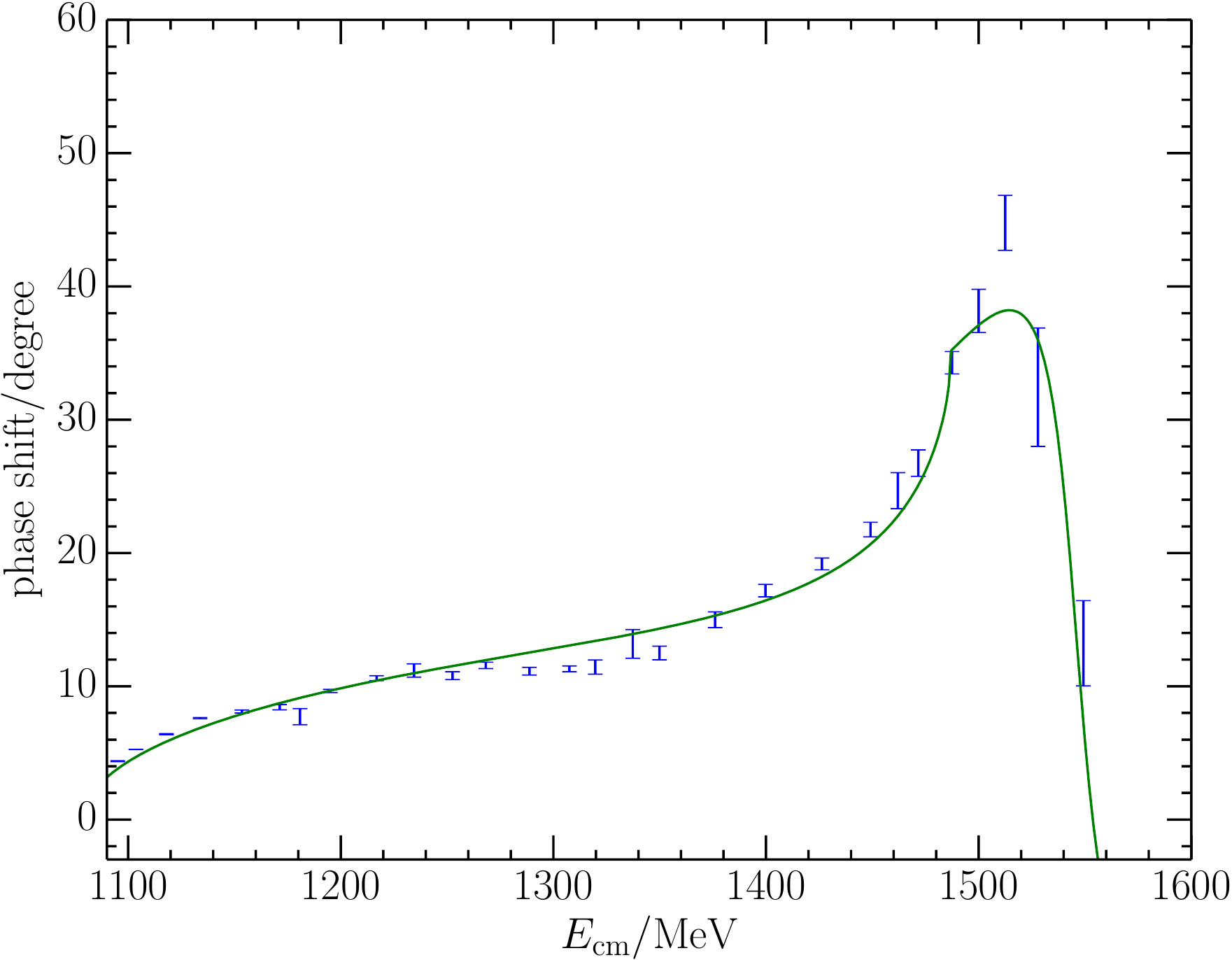}}}\qquad\quad
\subfigure{\scalebox{0.45}{\includegraphics{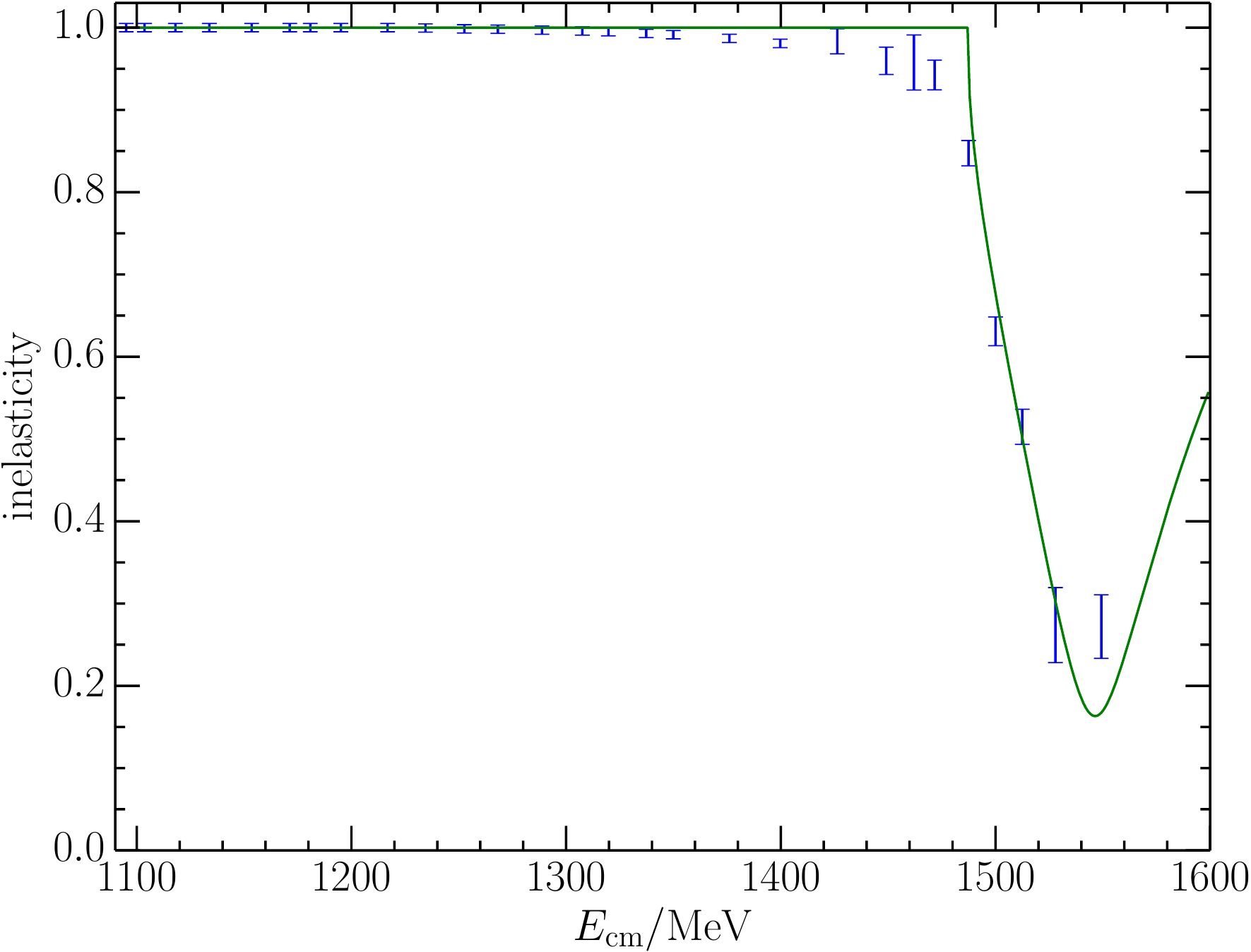}}}
\caption{{\bf Colour online:} Experimental data~\cite{NuclearStudies} for
the phase shift
  (left) and inelasticity (right) for $\pi N$ scattering with
  $J^P=1/2^-$ are fit by the Hamiltonian model.
  }
\label{fig:PSEta}
\end{center}
\end{figure*}

The scattering $T$ matrix is obtained from the relativistic
Lippmann-Schwinger equation.  The coupling parameters, $g_{N^*_0 \pi
  N},\ g_{N^*_0 \eta N}$, and $g^S_{\pi N}$, and the bare mass $m_0$
are determined by fitting the empirical phase shifts and
inelasticities for $\pi N$ scattering in the $J=1/2^-$ channel, with
guidance from the partial decay widths of the $N^*(1535)$
resonance.  
Varying these four parameters and fitting the 56 data points provides the fit
illustrated in Fig.~\ref{fig:PSEta}  with $\chi^2_{\rm dof}=6.8$ and
parameters:  
$g^S_{\pi N}=-0.0608\pm0.0004, \, 
g_{N^*_0 \pi N}=0.186\pm0.006,\, g_{N^*_0 \eta N}=0.185\pm0.017$ 
and $m_0=1601\pm14~{\rm MeV}$. This fit yields
a pole at  
$1531\pm 29 - i\, 88\pm 2$ MeV on 
the unphysical energy sheet for $\pi
N$ and $\eta N$, where the error of the pole only counts 
that of $m_0$.  This is in excellent agreement with the Particle
Data Group~\cite{Agashe:2014kda} estimate of $1510 \pm 20 - i\, 85 \pm
40$.
%

With the Hamiltonian model and associated parameters constrained by
experimental data, we can now calculate the
$J^P=1/2^-$ nucleon spectrum in the finite-volume considered in lattice QCD
calculations.
In a box with length $L$, the momentum a particle can carry in any
one dimension is constrained to integer multiples of the lowest
non-trivial momentum $2\pi/L$.  In three dimensions, it is convenient
to introduce the integer $n = n_x^2 + n_y^2 + n_z^2$ such that the
momenta available on the lattice are described by
$k_n=2\pi\sqrt{n}/L$. Full details of the translation of a 
Hamiltonian of the form given above into a Hamiltonian matrix 
on a finite spatial volume may be found in Ref.~\cite{Hall:2013qba}.

It has proven extremely useful in unravelling pieces of the strong 
interaction puzzle to move beyond the physical quark masses to the 
realm where they become larger. To explore this regime  
we allow the bare mass, 
$m_0$, to vary linearly with quark mass, so that (because $m_\pi^2
\sim m_q$), 
$m_0(m_\pi^2) \, = \, m_0|_{\rm
  phys}+\alpha_0(m_\pi^2-m_\pi^2|_{\rm phys})$. 
In the first instance, $\alpha_0$ is estimated through a
single parameter fit to current lattice QCD results for the $J^P=1/2^-$
nucleon spectrum.
The pion mass dependence of the ground state nucleon mass,
$m_N(m_\pi^2)$, is obtained via linear interpolation between the
lattice QCD results on the same size lattice.
The mass of the $\eta$ meson is related to the pion mass
via
$m_\eta^2(m_\pi^2) \, = \, m_\eta^2|_{\rm
  phys}+\frac{1}{3}(m_\pi^2-m_\pi^2|_{\rm phys})$.


\begin{figure*}[t]
\begin{center}
\scalebox{0.45}{\includegraphics{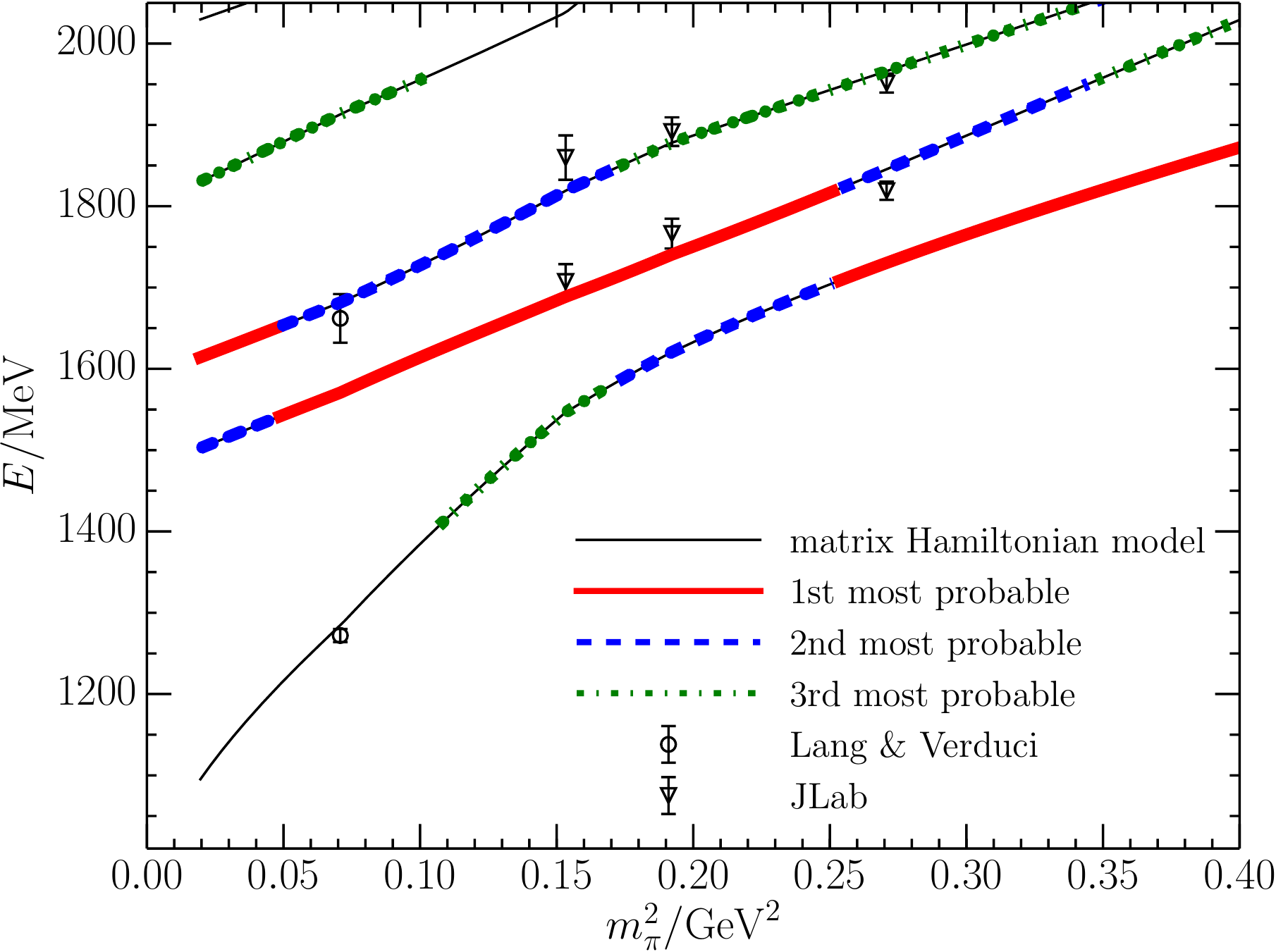}}\qquad\quad
\scalebox{0.45}{\includegraphics{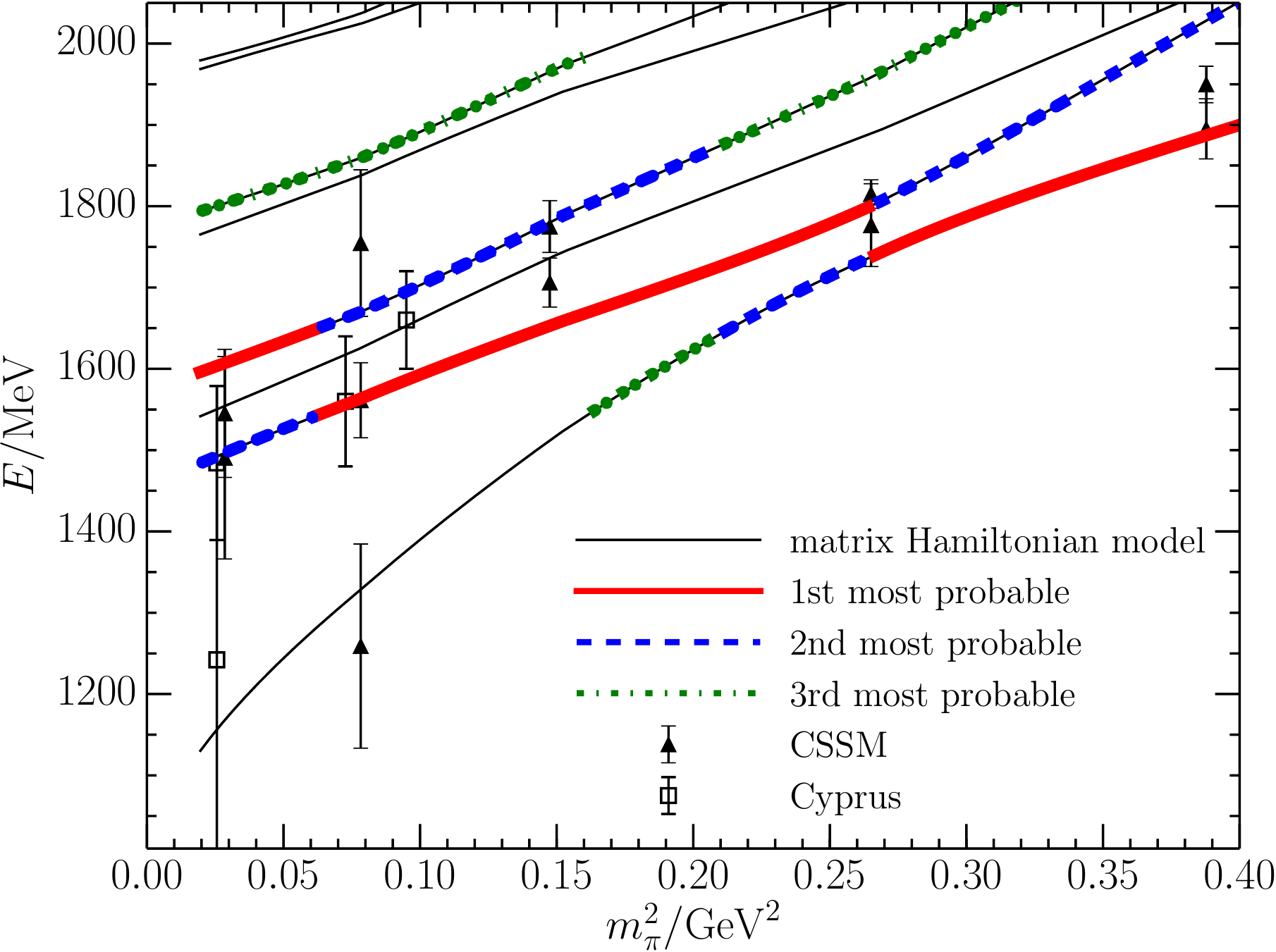}}
\caption{{\bf Colour online:} The pion mass dependence of the $L
  \simeq 1.98$ fm (left) and $L \simeq 2.90$ fm (right) finite-volume
  energy eigenstates.  The different line types and colours 
  indicate the strength of the bare
  basis state in the Hamiltonian-model eigenvector.
  }
\label{fig:Hamiltonian}
\end{center} 
\end{figure*}

The left-hand plot of Fig.~\ref{fig:Hamiltonian} illustrates results
from the Hadron Spectrum Collaboration~\cite{Edwards:2011jj, Edwards2013}
 (denoted JLab) and Lang
and Verduci~\cite{Lang:2012db}.  These precise results are obtained on
the smaller of the two lattice volumes considered herein, with length
$L \simeq 1.98$ fm.  The right-hand plot illustrates lattice QCD
results for lattice volumes with length $L\approx 2.90$ fm.  Recent
results from the
Centre for the Subatomic Structure of Matter (CSSM) lattice group in 
Adelaide~\cite{StokesEtAl,Kiratidis:2015vpa,Mahbub:2013ala,Mahbub:2012ri}
are shown, along with the Cyprus collaboration's results, obtained
using the Athens Model Independent Analysis Scheme
(AMIAS)~\cite{Alexandrou:2014mka}.
Both groups provide results for light pion masses $\simeq 160$ MeV.
The two lowest-lying odd-parity states from lattice QCD have an energy
similar to the non-interacting $S$-wave $\pi N$ scattering threshold.  
CSSM
reports two more low-lying states typically split by 100 MeV.  The
Cyprus collaboration reports one state in this regime with an energy
consistent with the lower of the two CSSM states.

The precision of the low-lying state observed by Lang and Verduci on
the 2 fm lattice highlights the different method employed in their
analysis.  There the low-lying scattering state was obtained by
creating a meson-baryon source in which the momentum of each hadron is
projected to zero.  In all other cases, the hadrons have been created
using conventional smeared-source operators.  To obtain the low-lying
state next to the non-interacting $S$-wave $\pi N$ scattering
threshold, the CSSM collaboration used five-quark operators.  All
other states have been obtained through the consideration of
three-quark operators.

In solving the matrix Hamiltonian the non-interacting basis states mix
to form eigenstates of the Hamiltonian.  These eigenstate energies are
illustrated in Fig.~\ref{fig:Hamiltonian} for lattice lengths $L
\simeq 1.98$ (left) and 2.90 fm (right).  Only one model parameter has
been adjusted in fitting 23 lattice energy eigenstates over three
levels on two volumes.  
The parameter
$\alpha_0 \, = \, 0.96\pm 0.06~{\rm GeV}^{-1}$, describing the 
quark-mass dependence of the bare
$N^*$ mass, was obtained from a simultaneous fit of these data
providing a $\chi^2_{\rm dof}=1.7$.
Of particular note is the excellent agreement between the
high-precision first state reported by Lang and Verduci
\cite{Lang:2012db} and the Hamiltonian model.  

Because the majority of the states observed in the lattice QCD
simulations have their origin in three-quark operators, we examine the
eigenvectors of the Hamiltonian states to identify states formed with
a large component of the bare basis state.  Under the assumption that
the three-quark operators couple most strongly to this bare state
component, one can then identify states in the matrix Hamiltonian
spectrum most likely associated with the states observed in the
lattice QCD simulations.  
In Fig.~\ref{fig:Hamiltonian} we have
indicated the strength of the bare-state component through different
line types and colours.  In both figures, the lattice QCD results
expected to be associated with resonant states are indeed described
well by the Hamiltonian model.  The Hamiltonian states dominated by
the bare-state component agree with the lattice results at the
one-standard-deviation level.

On comparing the Hamiltonian spectra presented in
Fig.~\ref{fig:Hamiltonian} one observes a significant dependence on
the volume of the lattice considered.  The additional complexity of
the spectrum encountered on larger lattice volumes is also apparent in
the right-hand plot of Fig.~\ref{fig:Hamiltonian}.  Here additional
meson-baryon dominated states appear next to the eigenstates seen on
the lattice.  Future simulations will include new meson-baryon
operators to capture these states in the lattice correlation-matrix
based variational analyses.

The lowest lying state on both lattice volumes is a $\pi N$ scattering
state at light quark masses, but this evolves into a state dominated
by the bare-mass component at heavy quark masses.  Here, as the mass
of the multi-particle state becomes very large, the lowest-lying state
is composed of a bare $N^*$ state dressed by $\pi N$ and $\eta N$
contributions.
The third eigenstate on the $L \simeq 2.90$ fm lattice which appears
between the two resonant-like lattice states is seen to be
predominately an $\eta N$ scattering state.

Next we turn to a more traditional analysis, 
where the aim is to 
extract information on the resonance of interest.
In this case, the low-energy
coefficients of the model, the bare mass $m_0$ and associated slope
$\alpha_0$, are both constrained by the lattice QCD results.  
After extracting these parameters from the fit to lattice data, 
we take the infinite volume limit and calculate the pole position.
In optimising these parameters, the Hamiltonian eigenstates dominated by
bare-state contributions are brought as close as possible to the
resonant-like lattice QCD results.  Similarly, the first state of the
Hamiltonian model is brought as close as possible to the lowest-lying
scattering states observed on the lattice.  A standard $\chi^2$
measure weighted by the lattice QCD energy uncertainties is used.  The
resultant fit is very good, with $\chi^2_{\rm dof} = 1.7$.  The main
change is a slight increase in the bare mass to better accommodate the
lattice QCD data at moderate pion masses.  Using a bootstrap analysis
to determine the standard errors from the percentiles of the
distributions, we find $m_0 = 1644\,_{-30}^{+34} {\rm~ MeV}$ and
$\alpha_0 = 0.77\,_{-0.16}^{+0.15}~{\rm GeV}^{-1}$, with the position of
the pole in the complex plane at: $1602 \pm 48 \, - \, i \, 
88.6\,_{-2.8}^{+0.7}~{\rm MeV}$.  The pole position lies just outside of
a one-sigma agreement with the Particle Data Group estimate of $1510
\pm 20 - i\, 85 \pm 40$.

The previous analysis included a background separable interaction 
which had been constrained by experimental data. 
Next, we explore the importance of such terms by dropping them 
and using only the
information provided by the lattice calculation. This is necessary, for
example, when there is insufficient experimental information on
its properties, especially its couplings to hadronic channels.
In particular, we 
fit the lattice QCD results by adjusting 
the two low-energy coefficients, $m_0$
and $\alpha_0$, but this time with the separable potential terms
discarded. The optimal fit yields a rather high  
$\chi^2_{\rm dof}=4.6$, largely because of the significant
discrepancy between the Hamiltonian model prediction for the lowest
lying $\pi N$ scattering state on the $L \simeq 1.98$ fm lattice and
the lattice QCD result of Lang and Verduci~\cite{Lang:2012db}.  
The majority of the resonant-like lattice results are still described
well by the Hamiltonian model.
Using a bootstrap analysis to obtain the uncertainties, 
the optimal parameters are
$m_0             = 1623\,_{-41}^{+33} {\rm~ MeV}, \, 
\alpha_0        =   0.85\,_{-0.17}^{+0.17}~{\rm GeV}^{-1}, \,
{\rm Re(pole)}  = 1563\,_{-80}^{+52}~{\rm MeV}, \,
{-\rm Im(pole)} =   89.2\,_{-4.2}^{+0.2}~{\rm MeV}$.
This pole position compares favorably with the Particle Data Group's
estimate of $1510 \pm 20 - i\, 85 \pm 40$.  However, the discrepancies
highlighted and the associated unacceptable $\chi^2_{\rm dof}=4.6$
reveals that the separable potential terms are vital 
to an accurate description of the lattice
QCD results.


In summary, we have used Hamiltonian effective field theory (HEFT) to study
the low-lying $J^P=1/2^-$ excitations of the nucleon in both the
finite volume of lattice QCD and the infinite volume of nature.
We have drawn on experimental data
for the lowest-lying $J^P=1/2^-$ nucleon resonance, the $N^*(1535)$,
and used HEFT to predict the positions of the finite-volume energy
levels to be observed in lattice QCD simulations in volumes of $\sim
2$ and $\sim 3$ fm.

The agreement between the HEFT predictions and lattice QCD results is
excellent and admits a more conventional analysis where the low-energy
coefficients are constrained by lattice QCD results, enabling a
determination of resonance properties from lattice QCD. We used
lattice QCD results from two different volumes to determine the pole
position of the $N^*(1535)$.  We find the pole position $1602 \pm 48 -
i\, 88.6\,_{-2.8}^{+0.7}~{\rm MeV}$, which lies just outside of
one-sigma agreement with the Particle Data Group estimate of $1510 \pm
20 - i\, 85 \pm 40$.

We also examined the role of $\pi N$ separable potential couplings and
found them to be essential in accurately describing the position of the
lowest-lying scattering state in the finite volume of the lattice.
The lattice length  dependence of the spectrum shows a rich structure
and it will be interesting to examine this in detail in future lattice
QCD calculations.

Finally, the success of this approach leads us to consider its
application to other $J^P$ baryon channels.  The $1/2^+$ channel of
the nucleon is of particular interest where evidence of the Roper
resonance in lattice QCD investigations is providing a fascinating
puzzle \cite{Leinweber:2015abc} waiting to be solved.  Application of
the successful formalism presented herein will be of benefit in
unravelling the mystery surrounding the Roper resonance in QCD.

\begin{acknowledgments}
This research is supported by the Australian Research Council through
the ARC Centre of Excellence for Particle Physics at the Terascale,
and through Grants DP120104627, DP150103164, DP140103067, LE120100181 (DBL) and 
FL0992247 and DP151103101 (AWT).
\end{acknowledgments}


\begin{thebibliography}{25}
\expandafter\ifx\csname natexlab\endcsname\relax\def\natexlab#1{#1}\fi
\expandafter\ifx\csname bibnamefont\endcsname\relax
  \def\bibnamefont#1{#1}\fi
\expandafter\ifx\csname bibfnamefont\endcsname\relax
  \def\bibfnamefont#1{#1}\fi
\expandafter\ifx\csname citenamefont\endcsname\relax
  \def\citenamefont#1{#1}\fi
\expandafter\ifx\csname url\endcsname\relax
  \def\url#1{\texttt{#1}}\fi
\expandafter\ifx\csname urlprefix\endcsname\relax\def\urlprefix{URL }\fi
\providecommand{\bibinfo}[2]{#2}
\providecommand{\eprint}[2][]{\url{#2}}

\bibitem[{\citenamefont{Luscher}(1986{\natexlab{a}})}]{Luscher:1985dn}
\bibinfo{author}{\bibfnamefont{M.}~\bibnamefont{Luscher}},
  \bibinfo{journal}{Commun. Math. Phys.} \textbf{\bibinfo{volume}{104}},
  \bibinfo{pages}{177} (\bibinfo{year}{1986}{\natexlab{a}}).

\bibitem[{\citenamefont{Luscher}(1986{\natexlab{b}})}]{Luscher:1986pf}
\bibinfo{author}{\bibfnamefont{M.}~\bibnamefont{Luscher}},
  \bibinfo{journal}{Commun. Math. Phys.} \textbf{\bibinfo{volume}{105}},
  \bibinfo{pages}{153} (\bibinfo{year}{1986}{\natexlab{b}}).

\bibitem[{\citenamefont{Wu et~al.}(2014)\citenamefont{Wu, Lee, Thomas, and
  Young}}]{Wu:2014vma}
\bibinfo{author}{\bibfnamefont{J.-J.} \bibnamefont{Wu}},
  \bibinfo{author}{\bibfnamefont{T.~S.~H.} \bibnamefont{Lee}},
  \bibinfo{author}{\bibfnamefont{A.~W.} \bibnamefont{Thomas}},
  \bibnamefont{and} \bibinfo{author}{\bibfnamefont{R.~D.} \bibnamefont{Young}},
  \bibinfo{journal}{Phys. Rev.} \textbf{\bibinfo{volume}{C90}},
  \bibinfo{pages}{055206} (\bibinfo{year}{2014}), \eprint{1402.4868}.

\bibitem[{\citenamefont{Hall et~al.}(2013)\citenamefont{Hall, Hsu, Leinweber,
  Thomas, and Young}}]{Hall:2013qba}
\bibinfo{author}{\bibfnamefont{J.~M.~M.} \bibnamefont{Hall}},
  \bibinfo{author}{\bibfnamefont{A.~C.~P.} \bibnamefont{Hsu}},
  \bibinfo{author}{\bibfnamefont{D.~B.} \bibnamefont{Leinweber}},
  \bibinfo{author}{\bibfnamefont{A.~W.} \bibnamefont{Thomas}},
  \bibnamefont{and} \bibinfo{author}{\bibfnamefont{R.~D.} \bibnamefont{Young}},
  \bibinfo{journal}{Phys. Rev.} \textbf{\bibinfo{volume}{D87}},
  \bibinfo{pages}{094510} (\bibinfo{year}{2013}), \eprint{1303.4157}.

\bibitem[{\citenamefont{Hall et~al.}(2015)\citenamefont{Hall, Kamleh,
  Leinweber, Menadue, Owen, Thomas, and Young}}]{Hall:2014uca}
\bibinfo{author}{\bibfnamefont{J.~M.~M.} \bibnamefont{Hall}},
  \bibinfo{author}{\bibfnamefont{W.}~\bibnamefont{Kamleh}},
  \bibinfo{author}{\bibfnamefont{D.~B.} \bibnamefont{Leinweber}},
  \bibinfo{author}{\bibfnamefont{B.~J.} \bibnamefont{Menadue}},
  \bibinfo{author}{\bibfnamefont{B.~J.} \bibnamefont{Owen}},
  \bibinfo{author}{\bibfnamefont{A.~W.} \bibnamefont{Thomas}},
  \bibnamefont{and} \bibinfo{author}{\bibfnamefont{R.~D.} \bibnamefont{Young}},
  \bibinfo{journal}{Phys. Rev. Lett.} \textbf{\bibinfo{volume}{114}},
  \bibinfo{pages}{132002} (\bibinfo{year}{2015}), \eprint{1411.3402}.

\bibitem[{\citenamefont{Nieves and Ruiz~Arriola}(2001)}]{Nieves2001}
\bibinfo{author}{\bibfnamefont{J.}~\bibnamefont{Nieves}} \bibnamefont{and}
  \bibinfo{author}{\bibfnamefont{E.}~\bibnamefont{Ruiz~Arriola}},
  \bibinfo{journal}{Phys. Rev. D} \textbf{\bibinfo{volume}{64}},
  \bibinfo{pages}{116008} (\bibinfo{year}{2001}).

\bibitem[{\citenamefont{Kaiser et~al.}(1995)\citenamefont{Kaiser, Siegel, and
  Weise}}]{Kaiser1995}
\bibinfo{author}{\bibfnamefont{N.}~\bibnamefont{Kaiser}},
  \bibinfo{author}{\bibfnamefont{P.}~\bibnamefont{Siegel}}, \bibnamefont{and}
  \bibinfo{author}{\bibfnamefont{W.}~\bibnamefont{Weise}},
  \bibinfo{journal}{Physics Letters B} \textbf{\bibinfo{volume}{362}},
  \bibinfo{pages}{23 } (\bibinfo{year}{1995}).

\bibitem[{\citenamefont{Inoue et~al.}(2002)\citenamefont{Inoue, Oset, and
  Vicente~Vacas}}]{Inoue2002}
\bibinfo{author}{\bibfnamefont{T.}~\bibnamefont{Inoue}},
  \bibinfo{author}{\bibfnamefont{E.}~\bibnamefont{Oset}}, \bibnamefont{and}
  \bibinfo{author}{\bibfnamefont{M.~J.} \bibnamefont{Vicente~Vacas}},
  \bibinfo{journal}{Phys. Rev. C} \textbf{\bibinfo{volume}{65}},
  \bibinfo{pages}{035204} (\bibinfo{year}{2002}).

\bibitem[{\citenamefont{Liu and Zou}(2006)}]{Liu2006}
\bibinfo{author}{\bibfnamefont{B.~C.} \bibnamefont{Liu}} \bibnamefont{and}
  \bibinfo{author}{\bibfnamefont{B.~S.} \bibnamefont{Zou}},
  \bibinfo{journal}{Phys. Rev. Lett.} \textbf{\bibinfo{volume}{96}},
  \bibinfo{pages}{042002} (\bibinfo{year}{2006}).

\bibitem[{\citenamefont{Olive et~al.}(2014)}]{Agashe:2014kda}
\bibinfo{author}{\bibfnamefont{K.~A.} \bibnamefont{Olive}} \bibnamefont{et~al.}
  (\bibinfo{collaboration}{Particle Data Group}), \bibinfo{journal}{Chin.
  Phys.} \textbf{\bibinfo{volume}{C38}}, \bibinfo{pages}{090001}
  (\bibinfo{year}{2014}).

\bibitem[{\citenamefont{Kamano et~al.}(2013)\citenamefont{Kamano, Nakamura,
  Lee, and Sato}}]{Kamano2013}
\bibinfo{author}{\bibfnamefont{H.}~\bibnamefont{Kamano}},
  \bibinfo{author}{\bibfnamefont{S.~X.} \bibnamefont{Nakamura}},
  \bibinfo{author}{\bibfnamefont{T.-S.~H.} \bibnamefont{Lee}},
  \bibnamefont{and} \bibinfo{author}{\bibfnamefont{T.}~\bibnamefont{Sato}},
  \bibinfo{journal}{Phys. Rev. C} \textbf{\bibinfo{volume}{88}},
  \bibinfo{pages}{035209} (\bibinfo{year}{2013}).

\bibitem[{\citenamefont{Matsuyama et~al.}(2007)\citenamefont{Matsuyama, Sato,
  and Lee}}]{Matsuyama2007}
\bibinfo{author}{\bibfnamefont{A.}~\bibnamefont{Matsuyama}},
  \bibinfo{author}{\bibfnamefont{T.}~\bibnamefont{Sato}}, \bibnamefont{and}
  \bibinfo{author}{\bibfnamefont{T.-S.} \bibnamefont{Lee}},
  \bibinfo{journal}{Physics Reports} \textbf{\bibinfo{volume}{439}},
  \bibinfo{pages}{193 } (\bibinfo{year}{2007}).

\bibitem[{\citenamefont{Kamano et~al.}(2011)\citenamefont{Kamano, Nakamura,
  Lee, and Sato}}]{Kamano2011}
\bibinfo{author}{\bibfnamefont{H.}~\bibnamefont{Kamano}},
  \bibinfo{author}{\bibfnamefont{S.~X.} \bibnamefont{Nakamura}},
  \bibinfo{author}{\bibfnamefont{T.-S.~H.} \bibnamefont{Lee}},
  \bibnamefont{and} \bibinfo{author}{\bibfnamefont{T.}~\bibnamefont{Sato}},
  \bibinfo{journal}{Phys. Rev. D} \textbf{\bibinfo{volume}{84}},
  \bibinfo{pages}{114019} (\bibinfo{year}{2011}).

\bibitem[{\citenamefont{{The Institute for Nuclear
  Studies}}(online)}]{NuclearStudies}
\bibinfo{author}{\bibnamefont{{The Institute for Nuclear Studies}}},
  \bibinfo{journal}{http://gwdac.phys.gwu.edu/}  (\bibinfo{year}{online}).

\bibitem[{\citenamefont{Edwards et~al.}(2011)\citenamefont{Edwards, Dudek,
  Richards, and Wallace}}]{Edwards:2011jj}
\bibinfo{author}{\bibfnamefont{R.~G.} \bibnamefont{Edwards}},
  \bibinfo{author}{\bibfnamefont{J.~J.} \bibnamefont{Dudek}},
  \bibinfo{author}{\bibfnamefont{D.~G.} \bibnamefont{Richards}},
  \bibnamefont{and} \bibinfo{author}{\bibfnamefont{S.~J.}
  \bibnamefont{Wallace}}, \bibinfo{journal}{Phys.Rev.}
  \textbf{\bibinfo{volume}{D84}}, \bibinfo{pages}{074508}
  (\bibinfo{year}{2011}), \eprint{1104.5152}.

\bibitem[{\citenamefont{Edwards et~al.}(2013)\citenamefont{Edwards, Mathur,
  Richards, and Wallace}}]{Edwards2013}
\bibinfo{author}{\bibfnamefont{R.~G.} \bibnamefont{Edwards}},
  \bibinfo{author}{\bibfnamefont{N.}~\bibnamefont{Mathur}},
  \bibinfo{author}{\bibfnamefont{D.~G.} \bibnamefont{Richards}},
  \bibnamefont{and} \bibinfo{author}{\bibfnamefont{S.~J.}
  \bibnamefont{Wallace}} (\bibinfo{collaboration}{Hadron Spectrum
  Collaboration}), \bibinfo{journal}{Phys. Rev. D}
  \textbf{\bibinfo{volume}{87}}, \bibinfo{pages}{054506}
  (\bibinfo{year}{2013}).

\bibitem[{\citenamefont{Lang and Verduci}(2013)}]{Lang:2012db}
\bibinfo{author}{\bibfnamefont{C.}~\bibnamefont{Lang}} \bibnamefont{and}
  \bibinfo{author}{\bibfnamefont{V.}~\bibnamefont{Verduci}},
  \bibinfo{journal}{Phys.Rev.} \textbf{\bibinfo{volume}{D87}},
  \bibinfo{pages}{054502} (\bibinfo{year}{2013}), \eprint{1212.5055}.

\bibitem[{\citenamefont{Stokes et~al.}()\citenamefont{Stokes, Kamleh, and
  Leinweber}}]{StokesEtAl}
\bibinfo{author}{\bibfnamefont{F.}~\bibnamefont{Stokes}},
  \bibinfo{author}{\bibfnamefont{W.}~\bibnamefont{Kamleh}}, \bibnamefont{and}
  \bibinfo{author}{\bibfnamefont{D.~B.} \bibnamefont{Leinweber}},
  \bibinfo{note}{in preparation.}

\bibitem[{\citenamefont{Kiratidis et~al.}(2015)\citenamefont{Kiratidis, Kamleh,
  Leinweber, and Owen}}]{Kiratidis:2015vpa}
\bibinfo{author}{\bibfnamefont{A.~L.} \bibnamefont{Kiratidis}},
  \bibinfo{author}{\bibfnamefont{W.}~\bibnamefont{Kamleh}},
  \bibinfo{author}{\bibfnamefont{D.~B.} \bibnamefont{Leinweber}},
  \bibnamefont{and} \bibinfo{author}{\bibfnamefont{B.~J.} \bibnamefont{Owen}},
  \bibinfo{journal}{Phys. Rev.} \textbf{\bibinfo{volume}{D91}},
  \bibinfo{pages}{094509} (\bibinfo{year}{2015}), \eprint{1501.07667}.

\bibitem[{\citenamefont{Mahbub et~al.}(2013{\natexlab{a}})\citenamefont{Mahbub,
  Kamleh, Leinweber, Moran, and Williams}}]{Mahbub:2013ala}
\bibinfo{author}{\bibfnamefont{M.~S.} \bibnamefont{Mahbub}},
  \bibinfo{author}{\bibfnamefont{W.}~\bibnamefont{Kamleh}},
  \bibinfo{author}{\bibfnamefont{D.~B.} \bibnamefont{Leinweber}},
  \bibinfo{author}{\bibfnamefont{P.~J.} \bibnamefont{Moran}}, \bibnamefont{and}
  \bibinfo{author}{\bibfnamefont{A.~G.} \bibnamefont{Williams}},
  \bibinfo{journal}{Phys. Rev.} \textbf{\bibinfo{volume}{D87}},
  \bibinfo{pages}{094506} (\bibinfo{year}{2013}{\natexlab{a}}),
  \eprint{1302.2987}.

\bibitem[{\citenamefont{Mahbub et~al.}(2013{\natexlab{b}})\citenamefont{Mahbub,
  Kamleh, Leinweber, Moran, and Williams}}]{Mahbub:2012ri}
\bibinfo{author}{\bibfnamefont{M.~S.} \bibnamefont{Mahbub}},
  \bibinfo{author}{\bibfnamefont{W.}~\bibnamefont{Kamleh}},
  \bibinfo{author}{\bibfnamefont{D.~B.} \bibnamefont{Leinweber}},
  \bibinfo{author}{\bibfnamefont{P.~J.} \bibnamefont{Moran}}, \bibnamefont{and}
  \bibinfo{author}{\bibfnamefont{A.~G.} \bibnamefont{Williams}},
  \bibinfo{journal}{Phys. Rev.} \textbf{\bibinfo{volume}{D87}},
  \bibinfo{pages}{011501} (\bibinfo{year}{2013}{\natexlab{b}}),
  \eprint{1209.0240}.

\bibitem[{\citenamefont{Alexandrou et~al.}(2015)\citenamefont{Alexandrou,
  Leontiou, Papanicolas, and Stiliaris}}]{Alexandrou:2014mka}
\bibinfo{author}{\bibfnamefont{C.}~\bibnamefont{Alexandrou}},
  \bibinfo{author}{\bibfnamefont{T.}~\bibnamefont{Leontiou}},
  \bibinfo{author}{\bibfnamefont{C.~N.} \bibnamefont{Papanicolas}},
  \bibnamefont{and}
  \bibinfo{author}{\bibfnamefont{E.}~\bibnamefont{Stiliaris}},
  \bibinfo{journal}{Phys. Rev.} \textbf{\bibinfo{volume}{D91}},
  \bibinfo{pages}{014506} (\bibinfo{year}{2015}), \eprint{1411.6765}.

\bibitem[{\citenamefont{Leinweber et~al.}(2015)\citenamefont{Leinweber, Kamleh,
  Kiratidis, Liu, Mahbub, Roberts, Stokes, Thomas, and Wu}}]{Leinweber:2015abc}
\bibinfo{author}{\bibfnamefont{D.}~\bibnamefont{Leinweber}},
  \bibinfo{author}{\bibfnamefont{W.}~\bibnamefont{Kamleh}},
  \bibinfo{author}{\bibfnamefont{A.}~\bibnamefont{Kiratidis}},
  \bibinfo{author}{\bibfnamefont{Z.-W.} \bibnamefont{Liu}},
  \bibinfo{author}{\bibfnamefont{S.}~\bibnamefont{Mahbub}},
  \bibinfo{author}{\bibfnamefont{D.}~\bibnamefont{Roberts}},
  \bibinfo{author}{\bibfnamefont{F.}~\bibnamefont{Stokes}},
  \bibinfo{author}{\bibfnamefont{A.~W.} \bibnamefont{Thomas}},
  \bibnamefont{and} \bibinfo{author}{\bibfnamefont{J.}~\bibnamefont{Wu}}
  (\bibinfo{year}{2015}), \eprint{1511.09146}.

\end{thebibliography}

\end{document}